\documentclass[useAMS,usenatbib,usegraphicx]{pasj00}
\usepackage{lscape}
\usepackage{color}
\usepackage{natbib, aas_macros}
\title{Universal Profiles of the Intracluster Medium from Suzaku X-Ray and Subaru Weak Lensing Obesrvations \thanks{Based on data collected at Subaru Telescope, which is operated by the National Astronomical Observatory of Japan.}}

 \author{%
N. \textsc{Okabe}, \altaffilmark{1}
K. \textsc{Umetsu}, \altaffilmark{2}
T. \textsc{Tamura}, \altaffilmark{3}
Y. \textsc{Fujita}, \altaffilmark{4}
M. \textsc{Takizawa}, \altaffilmark{5}
Y. -Y. \textsc{Zhang}, \altaffilmark{6}
K. \textsc{Matsushita}, \altaffilmark{7}
T. \textsc{Hamana}, \altaffilmark{8}
Y. \textsc{Fukazawa}, \altaffilmark{9}
T. \textsc{Futamase}, \altaffilmark{10}
M. \textsc{Kawaharada}, \altaffilmark{3}
S. \textsc{Miyazaki}, \altaffilmark{8}
Y. \textsc{Mochizuki}, \altaffilmark{7}
K. \textsc{Nakazawa}, \altaffilmark{11}
T. \textsc{Ohashi}, \altaffilmark{12}
N. \textsc{Ota}, \altaffilmark{13}
T. \textsc{Sasaki}, \altaffilmark{7}
K. \textsc{Sato}, \altaffilmark{7}
and
S. I. \textsc{Tam} \altaffilmark{14}}
\altaffiltext{1}{Kavli Institute for the Physics and Mathematics of the Universe (WPI), Todai Institutes for Advanced Study,University of Tokyo, 5-1-5 Kashiwanoha, Kashiwa, Chiba 277-8583, Japan}
 \email{nobuhiro.okabe@ipmu.jp}
\altaffiltext{2}{Institute of Astronomy and Astrophysics, Academia Sinica, P.O. Box 23-141, Taipei 10617, Taiwan}
\altaffiltext{3}{Institute of Space and Astronautical Science, Japan Aerospace Exploration Agency,3-1-1 Yoshinodai, Chuo-ku, Sagamihara, Kanagawa 229-8510, Japan}
\altaffiltext{4}{Department of Earth and Space Science, Graduate School of Science, Osaka University, Toyonaka, Osaka 560-0043, Japan}
\altaffiltext{5}{Department of Physics, Yamagata University, Kojirakawa-machi 1-4-12, Yamagata 990-8560, Japan}
\altaffiltext{6}{Argelander-Institut f\"ur Astronomie, Universit\"at Bonn, Auf dem H\"ugel 71, 53121 Bonn, Germany}
\altaffiltext{7}{Department of Physics, Tokyo University of Science, 1-3 Kagurazaka, Shinjyuku-ku, Tokyo 162-8601, Japan}
\altaffiltext{8}{National Astronomical Observatory of Japan, Mitaka, Tokyo 181-8588, Japan}
\altaffiltext{9}{Department of Physical Science, Hiroshima University, 1-3-1 Kagamiyama, Higashi-Hiroshima, Hiroshima 739-8526, Japan}
\altaffiltext{10}{Astronomical Institute, Tohoku University, Aramaki, Aoba-ku, Sendai 980-8578, Japan}
\altaffiltext{11}{Department of Physics, The University of Tokyo, 7-3-1 Hongo, Bunkyo-ku, Tokyo 113-0033, Japan}
\altaffiltext{12}{Department of Physics, Tokyo Metropolitan University, 1-1 Minami-Osawa, Hachioji, Tokyo 192-0397, Japan}
\altaffiltext{13}{Department of Physics, Nara Women's University, Kitauoyanishi-machi, Nara, Nara 630-8506, Japan}
\altaffiltext{14}{Department of Physics ,National Taiwan University, Taipei 10617, Taiwan}

\newcommand{\simgt}{\lower.5ex\hbox{$\; \buildrel > \over \sim \;$}}
\newcommand{\simlt}{\lower.5ex\hbox{$\; \buildrel < \over \sim \;$}}

\def\h70Msol{\mathrel{h_{70}^{-1}M_\odot}}

\begin{document}

\date{\today}


\maketitle

\label{firstpage}

\begin{abstract}
We conduct a joint X-ray and weak-lensing study of four relaxed galaxy
 clusters (Hydra A, A478, A1689 and A1835) observed by 
 both {\it Suzaku} and Subaru out to virial radii, with an aim to
 understand recently-discovered unexpected feature of
the intracluster medium (ICM) in cluster outskirts.
We show that the average hydrostatic-to-lensing total mass ratio for the four
 clusters decreases from $\sim 70\%$ to $\sim 40\%$ as the overdensity
 contrast 
 decreases from 500 to the virial value.
The average gas mass fraction from lensing total mass estimates 
 increases with cluster radius and agrees with the cosmic mean baryon
 fraction within the virial radius, whereas the X-ray-based gas fraction
considerably exceeds the cosmic values due to underestimation of the
 hydrostatic mass.
We also develop a new advanced method for determining normalized
 cluster radial profiles for multiple X-ray observables by simultaneously taking into
 account both their radial dependence and multivariate scaling relations with weak-lensing masses.
Although the four clusters span a range of halo mass, concentration, X-ray luminosity and redshift,
we find 
that the gas entropy, pressure, temperature and density profiles are all remarkably
 self-similar when scaled with the weak-lensing $M_{200}$ mass and
 $r_{200}$ radius.
The entropy monotonically increases out to $\sim 0.5r_{200}\sim r_{1000}$ 
following the accretion shock heating model $K(r)\propto r^{1.1}$, 
and flattens at $\simgt 0.5r_{200}$.
The universality of the scaled entropy profiles
indicates that the thermalization mechanism over the entire cluster region
 ($>0.1r_{200}$) is controlled by gravitation in a common way for all clusters, 
although
 the heating efficiency in the outskirts needs to be modified from the standard $r^{1.1}$ law.
The bivariate scaling functions of the gas density and temperature reveal 
that the flattening of the outskirts entropy profile is caused by the steepening of
 the temperature, rather than the flattening of the gas density.
\end{abstract}


\makeatletter

\section{Introduction}

Recent studies with the {\it Suzaku} X-ray satellite \citep{Mitsuda07} have
reported detections of very faint X-ray emission in the outskirts of
galaxy clusters thanks to its low and stable particle background.
These Suzaku observations
revealed unexpected observational features of the intracluster
medium (ICM) in the cluster outskirts
\cite[e.g.][]{Fujita08,Bautz09,Kawaharada10,Hoshino10,Simionescu11,Sato12,Walker12a,Walker12b,Walker13,Ichikawa13,Reiprich13}:
The observed gas temperature sharply declines beyond about half the
cluster virial radius. 
In the cluster outskirts, the gas temperature 
is at most 20-50\% of those at intermediate radii (from one-fourth to half the virial radius).  
The gas entropy profile $K(r)=k_B T(r)/n_e^{2/3}(r)$
flattens beyond about half the virial radius, 
in contrast to predictions of accretion shock-heating models
\citep[e.g.][]{Tozzi01,Ponman03}.
If all kinetic energy of the infalling gas is instantly thermalized by the accretion shock,
the entropy profile is predicted to increase with cluster radius as
$K(r)\propto r^{1.1}$ \citep{Tozzi01}, as supported by {\it Chandra} and
{\it XMM-Newton} observations of cluster central regions \citep[e.g.][]{Cavagnolo09,Pratt10}.
In fact, the
entropy in the interior region is enhanced compared to the $r^{1.1}$, as
also shown in \cite{Walker13}.
The outskirts entropy from {\it Suzaku} is systematically lower than 
these model predictions as well as those extrapolated from observations in the central regions.
These pieces of evidence indicate that the majority of electrons in the
cluster outskirts are not yet thermalized, so that
the thermal pressure in the outskirts 
is not sufficient to fully balance with the total gravity of the cluster.
Furthermore, \cite{Kawaharada10} and \cite{Ichikawa13} showed that the
anisotropic distributions of  gas temperature and entropy in the
outskirts are correlated with large-scale structure of galaxies outside
the central clusters.
\cite{Sato12} found that the gas density distribution is correlated with
large-scale structure of galaxies.
These spatial correlations between thermodynamic properties of the ICM
and large-scale environments indicate that the physical processes in the
cluster outskirts are influenced by surrounding cosmological environments.

Several interpretations have been proposed to explain the presence of
low entropy gas in the ICM.
For example,  hot ions could provide main thermal-pressure support \citep{Hoshino10} 
if the electron in the outskirts are not yet fully thermalized.
This could be possible because the timescale of thermal equilibrium of the electrons by Coulomb collisions is much longer than that of the ions. 
However, this scenario 
ignores the fact that the timescale for electrons to achieve thermal equilibrium 
governed by wave-particle interactions 
of plasma kinetic instabilities is much shorter than that by Coulomb
collisions. 
Alternatively, the kinetic energy of bulk and/or turbulent motions could
contribute to some fraction of the total pressure to  
fully balance with the gravity \citep{Kawaharada10}. 
\cite{Lapi10} and \cite{Cavaliere11} proposed that 
the outer slope of the potential becomes shallower due to the
acceleration of cosmic expansion and hence the efficiency of accretion
shock heating is weakened.
\cite{Fujita13b} proposed that the accretion of cosmic-rays at cluster
formation shocks consumes kinetic energy of infalling gas and  decreases
the entropy of the downstream gas. 
A high degree of gas clumpiness in cluster outskirts could lead to an overestimate of
the observed gas density, causing the apparent flattening of the derived
entropy profile \citep{Nagai11}. 
\citet{Simionescu11} found that the gas mass fraction in the Perseus
cluster based on hydrostatic mass estimation exceeds the cosmic baryon fraction within the virial radius,
and attributed this apparent excess to gas clumpiness.
However, their X-ray-only analysis suffers from
the inherent assumption of hydrostatic equilibrium.
In particular, in light of the observed low gas temperature and entropy
in cluster outskirts, 
the ICM there is expected to be out of hydrostatic equilibrium. 
Therefore, accurate and direct cluster mass determinations without the
hydrostatic equilibrium assumption  
are essential for understanding the physical state of the ICM in cluster outskirts.

Weak gravitational lensing techniques are complementary to X-ray observations
because weak-lensing mass estimates do not require any assumptions on
cluster dynamical states.
A coherent distortion pattern of background galaxy shapes 
caused by the gravitational potential of clusters
 enables us to recover the cluster mass distribution \citep[e.g.,][]{Bartelmann01}.
A comparison of X-ray and weak-lensing cluster mass estimates provides
 a stringent test of the level of hydrostatic equilibrium 
\citep[e.g.,][]{Zhang10,Kawaharada10,Mahdavi13}, 
which is important for empirically understanding the systematic bias in
cluster mass estimates and for constructing well-calibrated cluster
mass-observable relations for cluster cosmology 
\citep[e.g.][]{Vikhlinin09b}.
In particular, multi-wavelength cluster datasets of high quality covering the entire cluster region
are crucial for a diagnostic of the ICM states out to the virial radius.

In the present paper we compile a sample of four massive clusters (Hydra A, A478, A1689,
A1835) which have been deeply observed out to the virial radius
with the {\it Suzaku} X-ray satellite and the Subaru Telescope, and
perform a joint X-ray and weak-lensing analysis to study the relations
between the total mass and ICM in this cluster sample.
The four clusters have different properties of 
the X-ray luminosity ($L_X$), average temperature ($\langle k_B
T\rangle$) and redshift ($z$), as summarized in
Table \ref{tab:target}. 
The X-ray luminosites differ by one order of magnitude and the redshifts are at $0.05\simlt r\simlt0.25$.
The typical integration time with {\it Suzaku} to detect faint
X-ray emission from cluster outskirts is much longer than that for
cluster central regions and multi-pointing Subaru observations are
required to cover the entire region of nearby {\it Suzaku} targets. 
Therefore our current sample is limited only to four objects.
Weak-lensing and X-ray analyses are briefly described in Section \ref{sec:data}.
In Section \ref{sec:compare},
we compare our weak-lensing mass measurements with X-ray properties of
the ICM as function of the cluster overdensity radius.
The results are summarized
in Section \ref{sec:sum}.
In the paper, we use $\Omega_{m,0}=0.27$, $\Omega_{\Lambda}=0.73$ and $H_0=70h_{70}~{\rm km\,s^{-1}Mpc^{-1}}$.

\section{Data Analysis} \label{sec:data}

\subsection{Subaru lensing Analysis}

We carried out weak-lensing analyses of individual clusters in our
sample using wide-field multi-band observations taken with the Suprime-Cam
\citep{Miyazaki02} at the prime focus of the 8.2-m Subaru Telescope.
We securely selected background source galaxies in order to avoid
contamination by unlensed cluster galaxies.
We measure the mass $M_\Delta$  at overdensities of $\Delta=2500,1000,500,200$  
and the virial overdensity 
$\Delta_{\rm vir}\sim100-110$ \citep{Nakamura97}.
Here, $M_\Delta$ represents the mass enclosed within a sphere of radius $r_\Delta$ inside 
which the mean interior density is $\Delta$ times the critical mass
density, $\rho_{\rm cr}(z)$, at the redshift, $z$.

For A1689 ($z=0.1832$), we employ the nonparametrically-deprojected spherical mass model
from a joint strong-lensing, weak-lensing shear and
magnification analysis described in \cite{Kawaharada10} and \citet{Umetsu08}.
We fit the tangential (reduced) shear profile for the other three clusters with a parametrized mass
model. Here, the tangential distortion signal, the mean ellipticity of background galaxies
tangential to the cluster center, is obtained as a function of projected
distance from the brightest cluster galaxy (BCG).

Our weak-lensing analysis with new Subaru observations of Hydra A ($z=0.0538$) and A478 ($z=0.0881$) 
is described in details in \cite{Okabe14b}.
As for A1835 ($z=0.25280$), we have reanalyzed the Subaru data using new
background selection of \cite{Okabe13},  
and measured the mass profile in combination with strong lensing data \citep{Richard10}.
We employ the universal mass profile of \citet[hereafter NFW;][]{NFW96,NFW97} 
 which is empirically motivated by numerical simulations of
 collsionless cold dark matter.
The NFW density profile is given by the following form:
\begin{equation}
\rho_{\rm NFW}(r)=\frac{\rho_s}{(r/r_s)(1+r/r_s)^2},
\label{eq:rho_nfw}
\end{equation}
where $\rho_s$ is the central density parameter and $r_s$ is the scale radius.
The halo concentration is defined by $c_\Delta=r_\Delta/r_s$.
The resulting $M_{200}$ is listed in Table \ref{tab:target}.

Since the uncertainty of mass estimates by joint strong- and
weak-lensing measurements is much smaller than
those by weak-lensing-only measurements, we shall apply unweighted
averaging in our cluster ensemble analysis.

\begin{table*}
\caption{Target Properties. The X-ray luminosity is retrieved from the
 MCXC catalog \citep{Piffaretti11}. 
The fourth and fifth column are the average X-ray temperature and weak-lensing masses, respectively.
The references of the {\it Suzaku},{\it Chandra} and {\it XMM-Newton} data are listed.}\label{tab:target}
\begin{center}
\begin{tabular}{ccccccc}
\hline
\hline
Name     & $z$
     & $L_X$
     & $\langle k_B T \rangle$
     & $M_{200}$
     & {\it Suzaku}
     & {\it XMM-Newton}/{\it Chandra} \\
     &
     & [$10^{45}$ergs$^{-1}$]
     & [keV]
     & [$h_{70}^{-1}10^{14}M_\odot$]
     & 
     & \\
\hline
Hydra A & $0.0538$ 
       & $0.27$
       & $3.0$
       & $3.72_{-1.44}^{+2.11}$
       & \cite{Sato12} 
       & \cite{David01} \\
A 478  & $0.0881$
       & $0.72$
       & $7.0$
       & $13.05_{-3.30}^{+4.12}$
       & \cite{Mochizuki14} 
       & \cite{Sanderson05} \\
A 1689 & $0.1832$
       & $1.25$
       & $9.3$
       & $16.73_{-3.44}^{+4.88}$
       & \cite{Kawaharada10}
       & \cite{Zhang07} \\
A 1835 & $0.2528$
       & $1.97$
       & $8.0$
       & $10.35_{-2.40}^{+2.80}$
       & \cite{Ichikawa13}
       & \cite{Zhang07} \\
\hline
\end{tabular}
\end{center}
\end{table*}

\subsection{Suzaku X-ray Analysis}

The {\it Suzaku} studies for Hydra A, A478, A1689 and A1835 are described in details 
in \cite{Sato12}, \cite{Mochizuki14}, \cite{Kawaharada10} and \cite{Ichikawa13}, respectively.
We here briefly summarize the analyses.
Low and stable particle background of {\it Suzaku} 
is powerful to detect diffuse faint X-ray emission beyond about half of the virial radius.
On the other hand, high angular resolutions of {\it Chandra} and {\it XMM-Newton} 
have advantages to measure the ICM properties within about half of the virial radius.
A joint X-ray study, combined with these datasets in different sensitivities and resolutions, well constrains 
the temperature and density profiles from the cores out to virial radii \citep{Sato12,Ichikawa13,Mochizuki14}.
We also found that the X-ray observables (the density, temperature, pressure and entropy) 
derived by {\it Suzaku}, {\it XMM-Newton} and {\it Chandra} agree with each other 
at overlapping radii \citep{Zhang07,David01,Sanderson05}.
The X-ray surface brightness profile of {\it Suzaku} is consistent with the {\it ROAST} flux.
The thermal pressure out to virial radius measured by {\it Suzaku} is also in good agreement with
the {\it Planck} flux of the Sunyaev-Zel'dovich (SZ) effect on the
cosmic microwave background (CMB) \citep{Ichikawa13,Mochizuki14}.
We assume the spherical distribution to derive hydrostatic equilibrium masses and gas masses 
from best-fit functions of the temperature and gas density profiles, 
as described in our earlier papers \citep{Kawaharada10,Sato12,Ichikawa13,Mochizuki14}.
In the cluster outskirts, we use azimuthal averages of {\it Suzaku} X-ray observables for A478, A1689 and A1835.
As for Hydra A, we use unweighted averages of observables in two directions of 
an over-dense filamentary structure of galaxies and a low density void environment outside the cluster.
In the central regions, X-ray observables measured by the {\it XMM-Newton} or {\it Chandra} are added to the {\it Suzaku} data.
The references are listed in Table \ref{tab:target}. We take into account only statistical errors.

\section{Comparison of Lensing and X-ray measurements} \label{sec:compare}

\subsection{Comparison of Hydrostatic Equilibrium Mass and Weak-lensing Mass} \label{subsec:Mratio}

Weak-lensing mass measurements do not require the hydrostatic equilibrium assumption,
and are complementary to X-ray measurements.
It is of critical importance to compare these two independent mass
estimates for understanding the physical state of the ICM as well as for
examining the degree of hydrostatic equilibrium
\citep{Kawaharada10,Sato12,Ichikawa13,Mochizuki14}. 
Our unique dataset of wide-field {\it Suzaku} and Subaru observations
enables us to directly compare X-ray and lensing masses from the cluster core to
the virial radius.  

\begin{figure}
\includegraphics[width=\hsize]{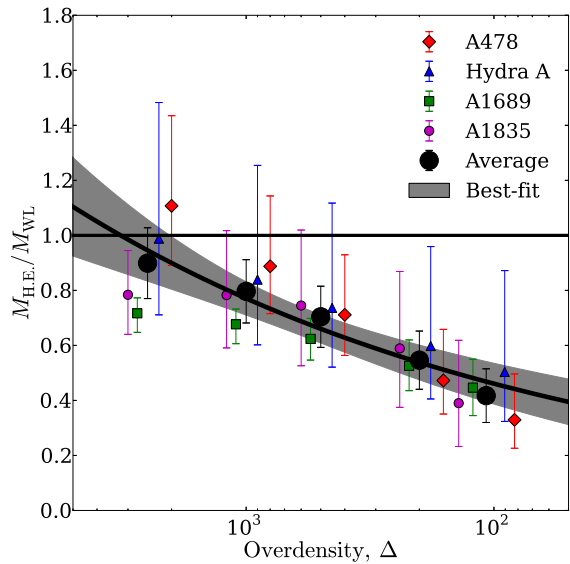}
\caption{X-ray hydrostatic to weak-lensing total mass ratios as a
 function of the density contrast $\Delta$.
Red diamonds, blue triangle, green squares, magenta circles and large black circles denote the mass ratios 
for A478, Hydra A, A1689, A1835 and the unweighted average of the 4 clusters, respectively.
From left to right, the data points with error bars represent the mass ratios at
 $\Delta=2500,1000,500,200$ and $\Delta_{\rm vir}$, respectively.
The mass ratios for the clusters are horizontally offset for
 visual clarity. 
The black-solid curve and gray-solid area are the best-fit profile and $1\sigma$ uncertainty.
}
\label{fig:Mratio_ave}
\end{figure}

Figure \ref{fig:Mratio_ave} shows the hydrostatic-to-lensing total mass ratio 
as a function of overdensity $\Delta$ 
with $\Delta=2500,1000,500,200$ and $\Delta_{\rm vir}$.
To avoid aperture-induced errors in enclosed-mass measurements, we
calculate the X-ray mass inside the same radius determined by lensing analysis.
Overall, the mass-ratio profiles decrease outward in a similar manner. 
We show in the figure that the unweighted average of the mass ratios
(large black circles) monotonically decreases as the 
overdensity $\Delta$ decreases.
We fit the average profile with the functional form of $\ln(\langle M_{\rm
H.E.}/M_{\rm WL}\rangle )=A+B\ln(\Delta)$. 
The best-fit slope is $B=0.22\pm0.07$. 
The overdensity dependence is thus detected at the $3\sigma$ level.
The mass discrepancy is negligible at a high overdensity of $\Delta=2500$.
However, we find that the X-ray hydrostatic mass can only account for $\sim70$\% and
$\sim40$\% of the lensing mass at $\Delta=500$ and $\Delta_{\rm vir}$, respectivly. 

Similar results were found inside $r_{500}$ by previous observational
studies and numerical simulations. 
\cite{Mahdavi13} compared their weak-lensing masses with X-ray masses
for 50 clusters,
finding that the X-ray to weak-lensing total mass ratio for their full sample is 
$0.92\pm0.05$ at $\Delta=2500$, $0.89\pm0.05$ at $\Delta=1000$,
and $0.88\pm0.05$ at $\Delta=500$, respectively,  
Hydrodynamical numerical simulations \citep{Nagai07,Piffaretti08,Lau09,Lau13,Nelson14} found 
that the hydrostatic-to-true total mass ratio within $r_{500}$ 
is more or less comparable with our results.
On the other hand, the hydrostatic-to-true mass ratio within $r_{200}$
is found to be $\sim80-90\%$, which is much larger than our results. 
Hence, there is a substantial discrepancy between our results and
numerical simulations in the cluster outskirts at low overdensities
$\Delta\simlt 200$.
Such a large bias can be caused if the hydrostatic cumulative mass unphysically 
 decreases with radius at cluster outskirts \citep[e.g.][]{Kawaharada10,Sato12}.  
We will discuss the possible deviations of X-ray observables in  Section \ref{subsec:joint}.

Measuring and quantifying any bias in hydrostatic mass estimates is one
of the key issues for cluster cosmology.
Lensing observations suffer from noisy projection effects, but can
provide unbiased cluster mass estimates in a statistical sense
if one can avoid an orientating bias \citep[e.g.,][]{Meneghetti2014}.
In particular, stacked cluster lensing measurements
\citep{Okabe10b,Okabe13,Oguri12,Umetsu11,Umetsu14},
which are insensitive to systematics due to projection effects, 
allow us to determine the representative mass profile for a cluster sample.​​

Our cluster sample is small but solely defined by the current availability of both {\it
Suzaku} and Subaru observations. 
We note that, although A1689 is a well-known strong-lensing cluster, the overall
trends in the observed mass ratios are common to all clusters.
Our results suggest 
that the degree of breakdown of the hydrostatic equilibrium assumption
can vary substantially with cluster radius,
indicating that an accurate characterization of the radial-dependent
mass bias is crucial for cluster cosmology.
This has direct consequences for the origin and degree of the apparent 
{\it tension} between the number counts of SZ clusters detected by {\it Planck} 
compared to those predicted by the {\it Planck} CMB cosmology
\citep{Planck13ClusterCosmology}. 
In their cluster cosmology analysis, \citet{Planck13ClusterCosmology}
adopted {\it XMM-Newton}-based hydrostatic mass estimates and calibrated
their cluster masses $M_{500}$ assuming a constant bias of 20\%.
Their assumed bias is smaller than but compatible with our
results.\footnote{Again, the X-ray mass has been measured within the
weak-lensing determined aperture radius in our study.}
Note that, since the size of our sample is small, it is essential to
conduct further systematic studies with larger samples
of X-ray and weak-lensing data covering the entire cluster region.

\subsection{Gas Fraction}

\begin{figure}
\includegraphics[width=\hsize]{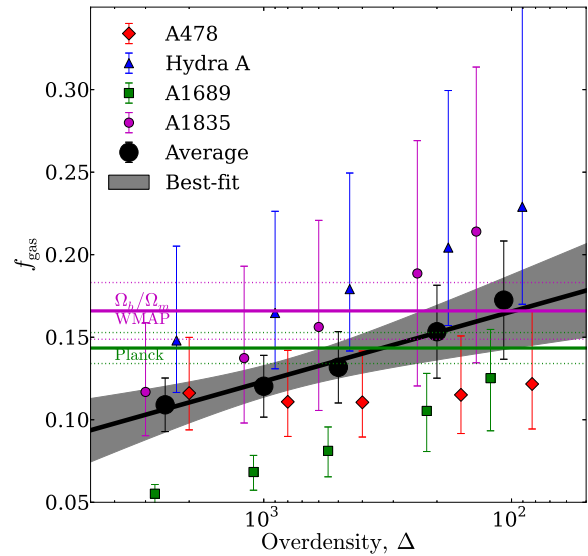}
\caption{Gas mass fraction, $M_{\rm gas}/M_{\rm WL}$, from weak-lensing
 total mass estimates, shown as a function of the density contrast $\Delta$.
Red diamonds, blue triangle, green squares, magenta circles and large black circles denote the mass ratios 
for A478, Hydra A, A1689, A1835 and the unweighted average of the 4 clusters, respectively.
The data points for the 4 clusters at each density are horizontally
 offset for visual clarity.
The horizontal-solid lines with dotted error bars are the cosmic mean
 baryon fractions from {\it WMAP} \citep{WMAP09} and {\it Planck}
 \citep{Planck13Cosmology} with their respective $1\sigma$ uncertainties.
The black-solid curve and gray-solid area show the best-fit profile
 and the $1\sigma$ uncertainty interval.}
\label{fig:fgas_ave}
\end{figure}

We have computed the gas mass fractions for the four clusters, 
$f_{\rm gas}(<r)=M_{\rm gas}(<r)/M_{\rm WL}(<r)$, as a function of the density
contrast $\Delta$ (Figure \ref{fig:fgas_ave}).
Here, the gas mass $M_{\rm gas}(<r)$ is measured inside 
the aperture radius $r_\Delta$ determined by weak-lensing analysis.
Compared to the total mass ratio, intrinsic scatter in gas fraction for
individual clusters is large.

Using the lensing total mass estimates, 
we find that the gas fractions within the virial radius are lower than or comparable to 
the cosmic mean baryon fraction \citep{WMAP09,Planck13Cosmology},
finding no evidence for the excess gas fraction relative to the cosmic value.

An apparent baryon excess within $r_{200}$ was reported in the Perseus
cluster on the basis of {\it Suzaku} hydrostatic mass estimates \citep{Simionescu11}.
However, as discussed in Section \ref{subsec:Mratio}, 
since the X-ray hydrostatic masses are underestimated especially at
large cluster radii,
the X-ray-based gas fractions can be largely overestimated and exceed
the cosmic mean baryon fraction
\citep[e.g.,][]{Sato12,Ichikawa13,Mochizuki14}. 
Weak-lensing mass determinations are needed to avoid such systematics.

Figure \ref{fig:fgas_ave} shows that the unweighted average of the gas
fractions increases as the overdensity decreases. 
We fit the average gas fraction profile with the functional form of
$\langle f_{\rm gas}\rangle=A+B\ln(\Delta)$.  
The best-fit normalization and slope are $A=0.250\pm0.065$ and $B=-0.018\pm0.009$, respectively.
The average gas fraction within the virial radius agrees within errors
with the cosmic mean baryon fractions from the {\it WMAP} \citep{WMAP09} 
and {\it Planck} \citep{Planck13Cosmology} experiments.

\subsection{Outskirts Entropy} \label{subsec:Kout}

The gas entropy profiles for relaxed clusters are observed to be
fairly universal over a wide radial range \citep{Walker12b,Sato12},
increasing following a power-law ($\propto r^{1.1}$) out to intermediate
radii and then flattening off from $\sim 0.5r_{\rm vir}$ to $r_{\rm vir}$.
Accordingly, the observed entropy in the outskirts is significantly lower
than that extrapolated with the power-law form of $K(r)\propto r^{1.1}$.
The scaled entropy profiles from {\it Suzaku} observations are well
fitted by a universal function in which the entropy flattening and
turnover are characterized by the virial radius or $r_{200}$
\citep{Walker12b,Sato12}, suggesting that there is a physical
correlation between the outskirts entropy and the virial mass.

\begin{figure*}
\includegraphics[width=\hsize]{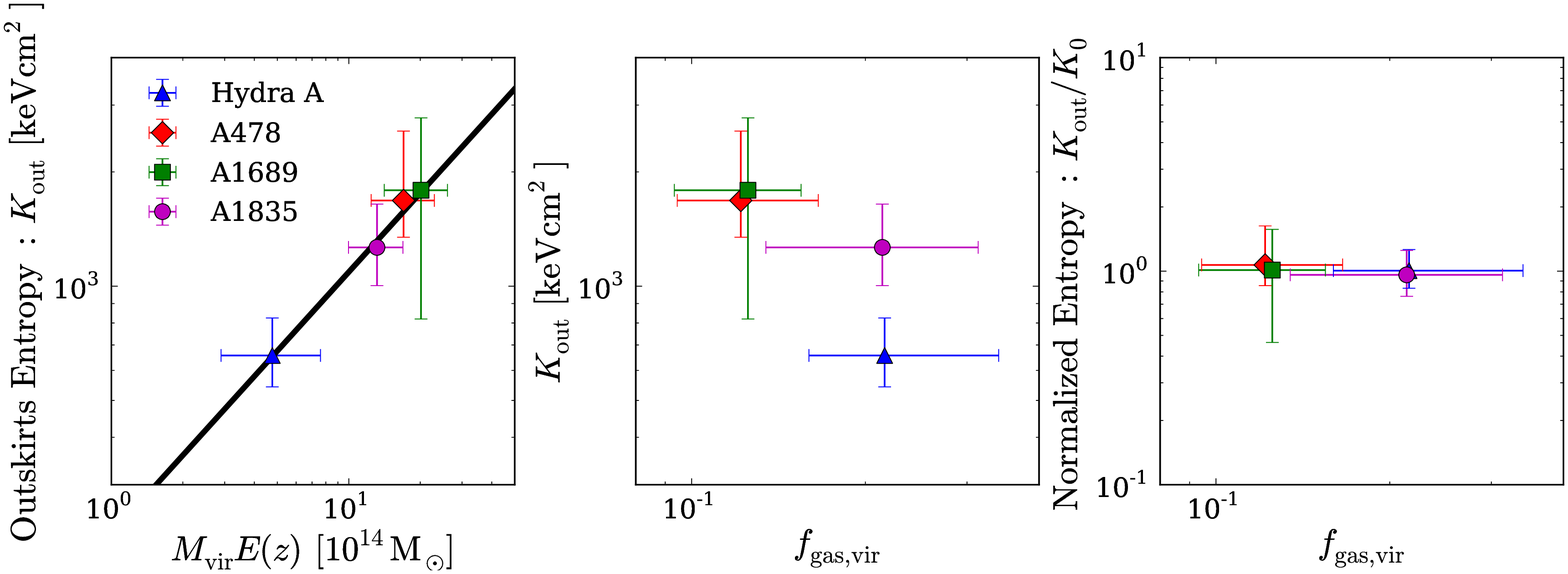}
\caption{Left: outskirts entropy ($K_{\rm out}$) versus virial mass
 ($M_{\rm vir}$) from lensing.
The correction factor $E(z)=H(z)/H_0$ accounts for the redshift
 evolution. Outskirts entropy is highly correlated with the virial
 mass. Middle: outskirts entropy versus gas fraction within the virial radius ($f_{\rm gas,vir}$),
Right: outskirts entropy normalized by the $K_{\rm out}-M_{\rm vir}$
 scaling relation ($K_0=K(M_{\rm vir})$) versus gas fraction. 
No significant correlation between the outskirts entropy and gas fractions is found.
}
\label{fig:K_ave}
\end{figure*}

Here we investigate 
a correlation between the outskirts entropy and the virial mass 
by using our lensing-derived virial
mass estimates.  
We adopt the average entropy $K(r)=k_B T/n_e^{2/3}$ in 
the radial range $r_{500}-r_{\rm vir}$ as the outskirts entropy, $K_{\rm out}$.
Since the observed entropy profiles in the outskirts are fairly flat, 
the results here are insensitive to the choice of the radial range. 
We fit the functional form $\ln(K_{\rm out})=A+B\ln(M_{\rm vir}E(z))$
for our sample of four clusters, 
where $E(z)=(\Omega_{m,0}(1+z)^3+\Omega_\Lambda)^{1/2}$ is the
dimensionless Hubble expansion rate, 
$A$ is the normalization and $B$ is the slope. 
The best-fit normalization and slope are $A=5.40\pm0.57$ and $B=0.69\pm0.25$, respectively.
We find a tight correlation between $K_{\rm out}$ and $E(z)M_{\rm vir}$
as shown in left panel of Figure \ref{fig:K_ave}.
Similarly, we also compare the relationship between the outskirts
entropy and $M_{200}$, finding again a tight correlation ($A=5.57\pm0.50$ and $B=0.69\pm0.24$).
Our results suggest that the gravity of the cluster has an important
effect on the thermalization process even in the outskirts.
However, since the outskirts entropy is lower than predicted by
the accretion shock heating model \citep{Tozzi01},  
the shock heating is not sufficient to explain the observations.

A possible mechanism for weakening of accretion shock heating in cluster
outskirts has been proposed by 
\cite{Lapi10} and \cite{Cavaliere11}.
According to their scenarios, the outer slope of the gravitational
potential becomes progressively shallower in the accelerating universe,
and the background gas density is decreased at late times accordingly.
Then the entropy production is reduced by the slowdown in the growth of
outskirts.
Their model predicts that the outskirts entropy is anti-correlated with
the halo concentration at a fixed mass because the concentration is
related to the formation epoch. 
The two clusters A478 and A1689 have similar virial masses, albeit different concentration parameters:
$c_{\rm vir}\sim4$ for the former \citep{Okabe14b} and $c_{\rm vir}\sim13$ for the latter \citep{Umetsu08}.
We do not find any significant difference in the outskirts entropy between the two clusters.
However, since the concentration parameter is sensitive to halo
triaxiality and orientation \citep[e.g.][]{Oguri04b} and the cluster
redshifts are different, the uncertainty from such other factors is large.
Therefore, a further systematic study with a large statistical sample is
required to investigate this possible correlation.

\cite{Nagai11} proposed that clumpy gas structures in cluster outskirts
lead to an overestimate of the gas density, so that 
the outskirts entropy can be underestimated from X-ray observations.
 If high gas clumpiness is a dominant source of the flat entropy, 
a correlation between the outskirts entropy and the gas fraction is expected.
However, in their simulations, gas clumpiness becomes dominant
only beyond $r_{200}$.
From {\it Suzaku} observations 
\cite{Simionescu11} found an extremely-high gas fraction within the
virial radius using their hydrostatic mass estimate for the Perseus cluster, 
and suggested that this is due to high gas clumpiness in the cluster outskirts.
Here we argue that the cumulative gas fraction used by \cite{Simionescu11} is not a
good quantity to discuss the degree of gas clumpiness which is locally defined.
Nevertheless, following \cite{Simionescu11}, 
we compare the outskirts entropy with the cumulative gas fractions
within the virial radii, in the middle panel of Figure \ref{fig:K_ave}.
Based on \cite{Simionescu11},     
it is expected that the gas fraction is anti-correlated with the outskirts entropy.
We find no clear correlation between $K_{\rm out}$ and $f_{\rm
gas,vir}\equiv f_{\rm gas}(<r_{\rm vir})$  for our three high-mass clusters.
Since there is a possibility that the scatter in $K_{\rm out}-f_{\rm gas,vir}$ plane is 
caused by the mass dependence on the entropy (left panel),
we show in the right panel a scaled version of the entropy normalized
using the scaling relation.
This shows that the scaled entropy does not correlate with the gas
fraction, indicating that one cannot attribute the high gas fraction and
low gas entropy to the gas clumping.

\subsection{Universal Entropy Profile} \label{subsec:K}

\citet{Walker12b} and \citet{Sato12} showed that the entropy profiles for relaxed
clusters have a universal form in the radial range from 
$\sim 0.1r_{200}$ to $\sim r_{200}$.
In their analyses, they first normalize entropy profiles to unity at a
certain pivot point (e.g., $0.3r_{200}$) and scale the cluster aperture
radii by $r_{200}$ which is determined according to the mass--temperature
relation.
Subsequently, fitting is performed to obtain the best-fit universal function.
However, such scaling operations make it difficult to correctly
propagate errors in the normalizations.
Since we use the weak-lensing mass $M_{\Delta}$ and the aperture radius
$r_{\Delta}$ for the normalizations,  their errors should be explicitly taken into account.  

To properly account for the errors in the weak-lensing and X-ray observables,
we avoid two-steps procedures of \citet{Walker12b} and \citet{Sato12}, and
perform a simultaneous ensemble fit of the observed entropy profiles as
a function of cluster radius and weak-lensing mass.
The log-likelihood function is defined by 
\begin{eqnarray}
-2\ln {\mathcal L}&=&\sum_{i,j}\ln(\delta_{\ln K,ij}^2+\delta_{\ln f,ij}^2+\sigma_{\ln K}^2) \nonumber \\
& &+ \frac{(\ln(K_i(r_j))-\ln(f_K(M_i,r_j)))^2}{\delta_{\ln K,ij}^2+\delta_{\ln f,ij}^2+\sigma_{\ln K}^2},
\end{eqnarray}
where $i$ and $j$ denote the $i$-th cluster and $j$-th radial bin, respectively.
$\delta_{\ln K}$ is the fractional error of the entropy,
$\delta_{\ln f}$ is the fractional error in the function $f_K$, through
its dependence on the total mass $M_{\Delta}$ and the aperture radius
$r_{\Delta}$ from weak-lensing, and
$\sigma_{\ln K}$ denotes intrinsic scatter of the entropy in the mass scaling relation.
Here we have introduced the function $f_K$, which takes into account the
flattening of the outskirts entropy profile, given by
\begin{eqnarray}
f_K(M_{\Delta},\tilde{r})&=&K_0E(z)^{-4/3}\left(\frac{M_{\Delta}E(z)}{10^{14}\h70Msol}\right)^{a} \nonumber \\
 &
  &\times(\tilde{r}/\tilde{r}_0)^\alpha\left(1+(\tilde{r}/\tilde{r}_0)^\beta\right)^{-\alpha/\beta}
  \label{eq:K}, 
\end{eqnarray}
where $K_0$ is the normalization factor, 
$\tilde{r}=r/r_{\Delta}$ is the aperture radius in units of $r_{\Delta}$, 
$a$ is the mass slope,
and $r_0$ denotes a characteristic scale radius at which the logarithmic
entropy gradient changes.
The asymptotic behavior is $K(r)\propto r^{\alpha}$ and $K(r)\propto
{\rm constant}$ for $r\ll r_0$ and $r\gg r_0$, respectively.
All errors in $M_{\Delta}$, $r_{\Delta}$ and $K$ are considered.

We perform fitting of two X-ray datasets.
The first dataset is composed only of the {\it Suzaku} data for the four clusters.
The second one includes the {\it XMM-Newton} and {\it Chandra} as well
as {\it Suzaku} data.
We restrict our X-ray data to $r>0.1r_{200}$ to excise the core regions.
The fitting is performed both with and without intrinsic scatter.

Since the {\it Suzaku} data alone cannot constrain the $\beta$
parameter,  we fix $\beta=4$ without intrinsic scatter. 
The resulting Bayesian estimates of the model parameters are listed in
Table \ref{tab:best-fit_K}. 
The results do not change significantly when we change $\beta$ by $\pm2$.
The inner slope of the entropy profile, $\alpha=1.18^{+0.93}_{-0.44}$,
agrees with the power-law slope ($K\propto r^{1.1}$) of the accretion
shock heating model \citep{Tozzi01}. 

\begin{figure}
\includegraphics[width=\hsize]{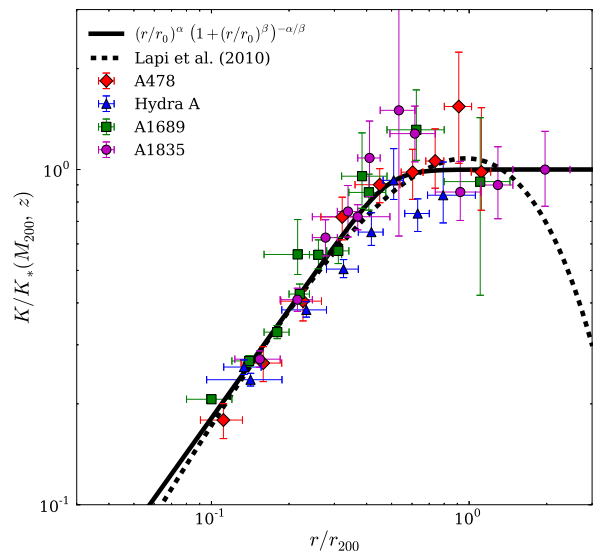}
\caption{Normalized entropy profile as a function of the radius scaled by $r_{200}$.
Red diamonds, blue triangle, green squares and magenta circles denote the normalized entropy for A478, Hydra A, A1689, and A1835, respectively. 
The normalization is $K_*(M_{200},z)=K_0E(z)^{-4/3}\left(M_{200}E(z)/10^{14}\h70Msol \right)^{a}$,
where $K_0=359.61\,[{\rm keV cm^2}]$ and $a=0.63$.
The black solid curve represents the best-fit universal entropy
 profile. The entropy profile at $r\simlt0.5r_{200}$ follows $r^{1.1}$ as
 predicted by the standard model \citep{Tozzi01} and becomes flat at  $r\simgt0.5r_{200}$.
The black-dashed curve shows the best-fit profile for the \citep{Lapi10} model.
}
\label{fig:K_prof}
\end{figure}

Next, we fit the full {\it Suzaku}, {\it XMM-Newton} and {\it Chandra}
data with and without intrinsic scatter. 
The constraint on the inner slope, $\alpha=1.11_{-0.13}^{+0.17}$, is
significantly improved.
The best-fit entropy function, which was obtained without
intrinsic scatter taken into account, gives an excellent description of
the scaled entropy profiles of the four clusters  (Figure
\ref{fig:K_prof}).
The mass dependence of the entropy is consistent with the self-similar model ($a=2/3$).
The inner radial slope $\alpha$ 
is in a remarkably good agreement with the shock heating model
\citep{Tozzi01} and with results from previous X-ray studies \citep[e.g.][]{Ponman03}.
The flattening of the entropy profile occurs at $r\sim 0.5r_{200}\sim
r_{1000}$, where the averaged hydrostatic to weak-lensing mass ratio is
less than unity (Section \ref{subsec:Mratio}).

We note that the four clusters studied here span a range of halo mass,
concentration, X-ray luminosity and redshift, but exhibit a remarkable
self-similarity in the scaled entropy profiles over the entire cluster ($>0.1r_{200}$).
In other words, not only the $r^{1.1}$-law's entropy profile but also the outskirts flattening entropy
depends on cluster $M_{200}$ masses and $r_{200}$ radius.
The results indicate that
the thermalization process in the ICM outside X-ray cores could be governed by the gravity of the cluster. 
The efficiency of gravitational heating from outside the core to
$0.5r_{200}$ follows the accretion shock heating model \citep{Tozzi01}, 
whereas the efficiency at $r\simgt0.5r_{200}$ is lower than the model
prediction.
The universality, independent of cluster properties, indicates that
the thermalization mechanism at
work in the ICM could be a common physical process to all clusters, perhaps
controlled by the growth of large sale structure surrounding the
cluster, although the heating efficiency in the outskirts may need to be
modified from the standard shock-heating mechanism \citep{Tozzi01}.

We also fit our data with an alternative functional form of $K\propto \tilde{r}^A\exp(B(1-\tilde{r}))$
proposed by \cite{Lapi10} and \cite{Cavaliere11}, by including our
 normalization parameter and using $\tilde{r}=r/r_{\rm vir}$.
We find this model also gives a good fit to the data.  To statistically
distinguish these two models, we need a larger sample of cluster
observations with a wide radial coverage beyond the virial radius.

\begin{table*}
\caption{Best-fit parameters for the entropy profile. The results of the
 full {\it Suzaku}$+${\it XMM-Newton}$+${\it Chandra} data is referred
 to as ``Full''.} \label{tab:best-fit_K} 
\begin{center}
\begin{tabular}{cccccccc}
\hline
\hline
Dataset     & $\Delta$
     & $K_0$
     & $a$
     & $\alpha$
     & $\beta$
     & $r_0$ 
     & $\sigma_{\ln K}$\\
     &
     & [keV cm$^2$]
     & 
     & 
     & 
     & [$r_{\Delta}$]
     &  \\
\hline
\it{Suzaku} & $200$
     & $283.52_{-80.71}^{+102.98}$
     & $0.71_{-0.11}^{+0.14}$
     & $1.18_{-0.44}^{+0.93}$
     & $4$ (fixed)
     & $0.41_{-0.16}^{+0.25}$
     & $-$  \\
Full & $200$
     & $359.61_{-75.47}^{+92.00}$
     & $0.63_{-0.08}^{+0.09}$
     & $1.11_{-0.13}^{+0.17}$
     & $5.97_{-2.87}^{+2.49}$
     & $0.48_{-0.08}^{+0.10}$
     & $-$  \\
Full & $200$
     & $380.49_{-71.61}^{+81.79}$
     & $0.62_{-0.07}^{+0.08}$
     & $1.10_{-0.13}^{+0.15}$
     & $5.97$ (fixed)
     & $0.49_{-0.07}^{+0.10}$
     & $<0.06$  \\
\hline
\end{tabular}
\end{center}
\end{table*}

\subsection{Universal Pressure Profile} \label{subsec:P}

It was shown by \cite{PlanckSZPprof} that the stacked pressure profile $P_e=n_ek_BT$
constructed from {\it Planck} SZ observations of 62 clusters
is well described by a generalized NFW pressure profile \citep{Nagai07b} out to $3\times r_{500}$. 
It is of great importance to compare our independent {\it Suzaku} X-ray pressure
measurements to the {\it Planck} SZ observations \citep{Walker12a}.

Here, we derive the average electron pressure profile scaled with the
weak-lensing mass $M_\Delta$ following the same procedure described in
Section \ref{subsec:K}.  
We consider the universal pressure function
\citep{Nagai07b,PlanckSZPprof} of the following form, by simultaneously
taking into account the scaling relation between the 
electron pressure and total
mass $M_\Delta$:
\begin{eqnarray}
f_P(M_{\Delta},\tilde{r})&=&P_0E(z)^{2}\left(\frac{M_{\Delta}E(z)}{10^{14}\h70Msol}\right)^{b} \nonumber \\
& &\times (\tilde{r}/\tilde{r}_0)^{-\gamma}\left(1+(\tilde{r}/\tilde{r}_0)^\beta\right)^{(\gamma-\delta)/\beta}, \label{eq:P}
\end{eqnarray}
where $P_0$ is the normalization factor, $\tilde{r}=r/r_{\Delta}$ is the
aperture radius in units of $r_{\Delta}$, 
and the mass slope $b$ accounts for the dependence of the pressure
normalization on
the halo mass $M_\Delta$ determined from the lensing analysis.
The asymptotic entropy slopes are $P\propto r^{-\gamma}$ and $r^{-\delta}$ for $r\ll r_0$ and $r\gg r_0$, 
respectively. The inverse of the scale radius, ${\tilde r}_0^{-1}=r_{\Delta}/r_0$, 
is equivalent to the concentration parameter, $c_\Delta$, in the definition of \cite{PlanckSZPprof}.

\begin{figure}
\includegraphics[width=\hsize]{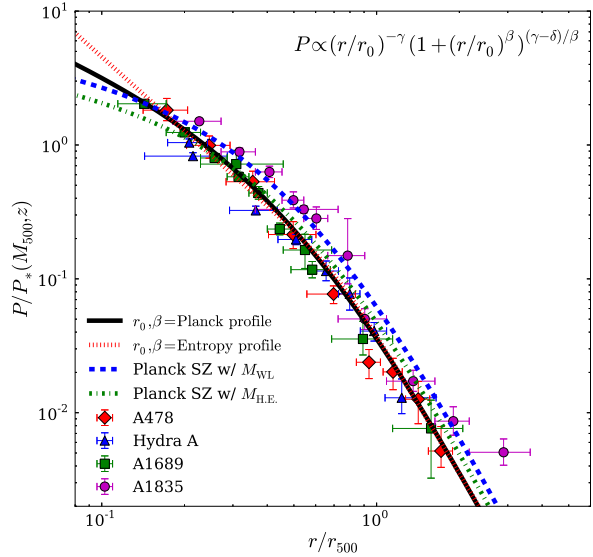}
\caption{Normalized pressure profile as a function of cluster radius scaled by $r_{500}$. 
Red diamonds, blue triangle, green squares and magenta circles represent
 the normalized pressure for A478, Hydra A, A1689, and A1835,
 respectively. The normalization is
 $P_*(M_{500},z)=P_0E(z)^{1/2}\left(M_{500}E(z)/10^{14}\h70Msol
 \right)^{b}$. 
The black-solid and red-dotted lines are the best-fit profiles using $r_0$
 and $\beta$ from the {\it Planck} results \citep{PlanckSZPprof} and our
 best-fit entropy profile,
 respectively. The blue-dashed and green-dotted-dashed lines are the
 average SZ pressure profiles \citep{PlanckSZPprof}, obtained using the
 weak-lensing and hydrostatic masses, respectively.
}
\label{fig:P_prof}
\end{figure}

First, we fit our pressure profiles at the reference overdensity 
$\Delta=500$ to make a fair comparison with \cite{PlanckSZPprof}. 
 We fix $r_0$ and $\beta$ by the {\it Planck} results ($c_{500}={\tilde
 r}_{0}^{-1}=1.81$ and $\beta=1.33$), because it is difficult to constrain them from our data.
The best-fit parameters are listed in Table \ref{tab:best-fit_P}. 
The normalized pressure profiles derived from the full {\it Suzaku},
{\it XMM-Newton} and {\it Chandra} dataset
 are displayed in Figure \ref{fig:P_prof}. 
The logarithmic gradient of the pressure profile progressively steepens
from $-1$ to $-4$. 

For comparison, we calculate the normalized {\it Planck} pressure
profiles using the equations (7), (10) and (11) of \cite{PlanckSZPprof}.
We use the best-fit parameters of the stacked pressure profile 
and determine the normalization by two approaches, 
namely using our weak-lensing or X-ray hydrostatic mass estimates,
instead of their original mass estimates using the $Y_X=M_{\rm gas}T_X$
mass proxy.
Figure \ref{fig:P_prof} shows that
the {\it Planck} average pressure profiles agree with the {\it Suzaku}
results.
The outer slope $\delta$ is in excellent agreement with the average
slope of the {\it Planck} SZ pressure profile ($\delta=4.1$). 
Although our sample size is much smaller than that of the {\it Planck} sample,
our results show good consistency between the observationally
independent X-ray and SZ profiles of individual clusters
\citep{Walker12a,Mochizuki14}.

Next, we fix $r_0=0.48r_{200}$ and $\beta=5.97$ using the best-fit entropy
profile (Section \ref{subsec:K}).  Now we derive scaled profiles at a
density contrast of $\Delta=200$.
The best-fit pressure profile, converted to $r_{500}$ units, is shown in Figure \ref{fig:P_prof}.
The resulting pressure profile is compatible with our and the average
{\it Planck} pressure profiles.
We also measure intrinsic scatter with $b=2/3$ fixed (Table \ref{tab:best-fit_P}).
Some parameters, including $b$, $\gamma$, and $\delta$,
are slightly changed depending on the choice of $\beta$ and $r_0$. 
A more definitive determination of these parameters requires a further systematic study.

\begin{table*}
\caption{Best-fit parameters for the pressure profile. 
``Full'' refers to  the full {\it Suzaku}$+${\it XMM-Newton}$+${\it Chandra} dataset.} \label{tab:best-fit_P} 
\begin{center}
\begin{tabular}{ccccccccc}
\hline
\hline
Dataset     & $\Delta$
     & $\log_{10}(P_0)$
     & $b$
     & $\beta$
     & $\gamma$
     & $\delta$
     & $r_0$
     & $\sigma_{\ln P}$ \\
     &
     & [log(keV cm$^{-3}$)]
     & 
     & 
     & 
     & 
     & [$r_{\Delta}$]
     & \\
\hline
\it{Suzaku} & $500$
     & $-2.41_{-0.17}^{+0.17}$
     & $0.43_{-0.10}^{+0.10}$
     & $1.33$ (fixed)
     & $1.08_{-0.25}^{+0.25}$
     & $3.41_{-0.38}^{+0.42}$
     & $1/1.81$ (fixed)
     & $-$ \\
Full & $500$
     & $-2.00_{-0.14}^{+0.15}$
     & $0.08_{-0.05}^{+0.07}$
     & $1.33$ (fixed)
     & $0.81_{-0.21}^{+0.22}$
     & $4.07_{-0.35}^{+0.36}$
     & $1/1.81$ (fixed)
     & $-$ \\
Full & $500$
     & $-2.35_{-0.21}^{+0.2}$
     & $0.15_{-0.08}^{+0.10}$
     & $1.33$ (fixed)
     & $1.24_{-0.38}^{+0.38}$
     & $4.07_{-0.35}^{+0.36}$
     & $1/1.81$ (fixed)
     & $0.24_{-0.06}^{+0.10}$ \\
Full & $200$
     & $-3.14_{-0.16}^{+0.16}$
     & $0.32_{-0.14}^{+0.14}$
     & $5.97$ (fixed)
     & $1.84_{-0.15}^{+0.17}$
     & $3.45_{-0.36}^{+0.43}$
     & $0.48$ (fixed)
     & $-$\\
Full & $200$
     & $-2.85_{-0.36}^{+0.38}$
     & $2/3$  (fixed)
     & $5.54_{-2.84}^{+2.81}$
     & $1.26_{-0.79}^{+0.44}$
     & $2.67_{-0.21}^{+0.30}$
     & $0.24_{-0.07}^{+0.11}$
     & $0.28_{-0.09}^{+0.10}$\\
\hline
\end{tabular}
\end{center}
\end{table*}

\subsection{Joint Fit with the Number Density and Temperature Profiles}\label{subsec:joint}

In Sections \ref{subsec:K} and \ref{subsec:P}, we have shown that the scaled
entropy and pressure profiles of the four clusters 
are well fitted with respective universal
functions, by taking into account the uncertainties in
the weak-lensing mass and X-ray measurements.

Here we simultaneously fit the gas density and temperature profiles to
our X-ray observations for determining the respective universal functions.
From equations (\ref{eq:K}) and (\ref{eq:P}), 
we obtain the following expressions for the gas number density and temperature:
\begin{eqnarray}
f_n &=&n_0E(z)^{2}\left(\frac{M_{\Delta}E(z)}{10^{14}\h70Msol}\right)^{\frac{3}{5}(b-a)}\nonumber \\
&& \times(\tilde{r}/\tilde{r}_0)^{-\frac{3}{5}(\alpha+\gamma)}\left(1+(\tilde{r}/\tilde{r}_0)^\beta\right)^{-\frac{3}{5}(\delta-\gamma-\alpha)/\beta},\label{eq:n}\\
f_T&=&T_0\left(\frac{M_{\Delta}E(z)}{10^{14}\h70Msol}\right)^{\frac{3}{5}a+\frac{2}{5}b} \nonumber \\
&& \times(\tilde{r}/\tilde{r}_0)^{\frac{3}{5}\alpha-\frac{2}{5}\gamma}\left(1+(\tilde{r}/\tilde{r}_0)^\beta\right)^{-(\frac{2}{5}\delta-\frac{2}{5}\gamma+\frac{3}{5}\alpha)/\beta}, \label{eq:T}
\end{eqnarray}
where $M_{\Delta}$ is the total cluster mass from lensing measurements, 
$n_0$ and $T_0$ are the normalization factors for the gas density and
temperature profiles, respectively.
Here we have assumed that the gas density and temperature profiles have
the same scale radius $r_0$ and $\beta$.

The joint likelihood for the number density and the temperature profiles
is given by 
\begin{eqnarray}
-2\ln {\mathcal L}&=&\sum_{i,j}\ln(\det(\mbox{\boldmath $C$}_{ij})) + \mbox{\boldmath $v$}_{ij}^T\mbox{\boldmath $C$}_{ij}^{-1}\mbox{\boldmath $v$}_{ij},  \\
\mbox{\boldmath $v$}&=&\left(
\begin{array}{ccc}
\ln(n(\tilde{r}))-\ln(f_n(M_{\Delta},\tilde{r})) \\
\ln(T(\tilde{r}))-\ln(f_T(M_{\Delta},\tilde{r})) \\
\end{array}
\right), \nonumber \\
\mbox{\boldmath $C$}&=&\mbox{\boldmath $C$}_{\rm stat}+\mbox{\boldmath $C$}_{\rm int} \nonumber \\
\mbox{\boldmath $C$}_{\rm stat}&=&\left(
\begin{array}{ccc}
\delta_{\ln n}^2+\delta_{\ln fn}^2 & \delta_{\ln fn,\ln fT} \\
\delta_{\ln fn,\ln fT} & \delta_{\ln T}^2+\delta_{\ln fT}^2 \\
\end{array}
\right). \nonumber \\
\mbox{\boldmath $C$}_{\rm int}&=&\left(
\begin{array}{ccc}
\sigma_{\ln n}^2 & \rho\sigma_{\ln n}\sigma_{\ln T} \\
\rho\sigma_{\ln n}\sigma_{\ln T} & \sigma_{\ln T}^2 \\
\end{array}
\right). \nonumber
\end{eqnarray}
Here, $i$ and $j$ denote the $i$-th cluster and $j$-th radial bins, respectively.
$\mbox{\boldmath $C$}_{\rm stat}$ is the error covariance matrix for the
data vector \mbox{\boldmath $v$},
$\delta_{\ln n}$ and $\delta_{\ln T}$ are the fractional errors of the
gas density and temperature,
$\sigma_{\ln f_n}$ and $\sigma_{\ln f_T}$ are the fractional errors in
the functions $f_n$ and $f_T$ (see equations (\ref{eq:n}) and (\ref{eq:T}))
through their weak-lensing mass dependence.
We assume that the error correlation between the number density and the temperature is negligible.
Note that the off-diagonal elements in the covariance matrix cannot be
ignored because both the gas density and temperature depend on the weak-lensing mass.
The intrinsic covariance $\mbox{\boldmath $C$}_{\rm int}$ 
consists of intrinsic scatter of the number density $\sigma_{\ln n}$ and
the temperature $\sigma_{\ln T}$.
By definition the correlation coefficient $\rho$ is in the range $-1\leq \rho\leq 1$.

\begin{figure*}
\includegraphics[width=\hsize]{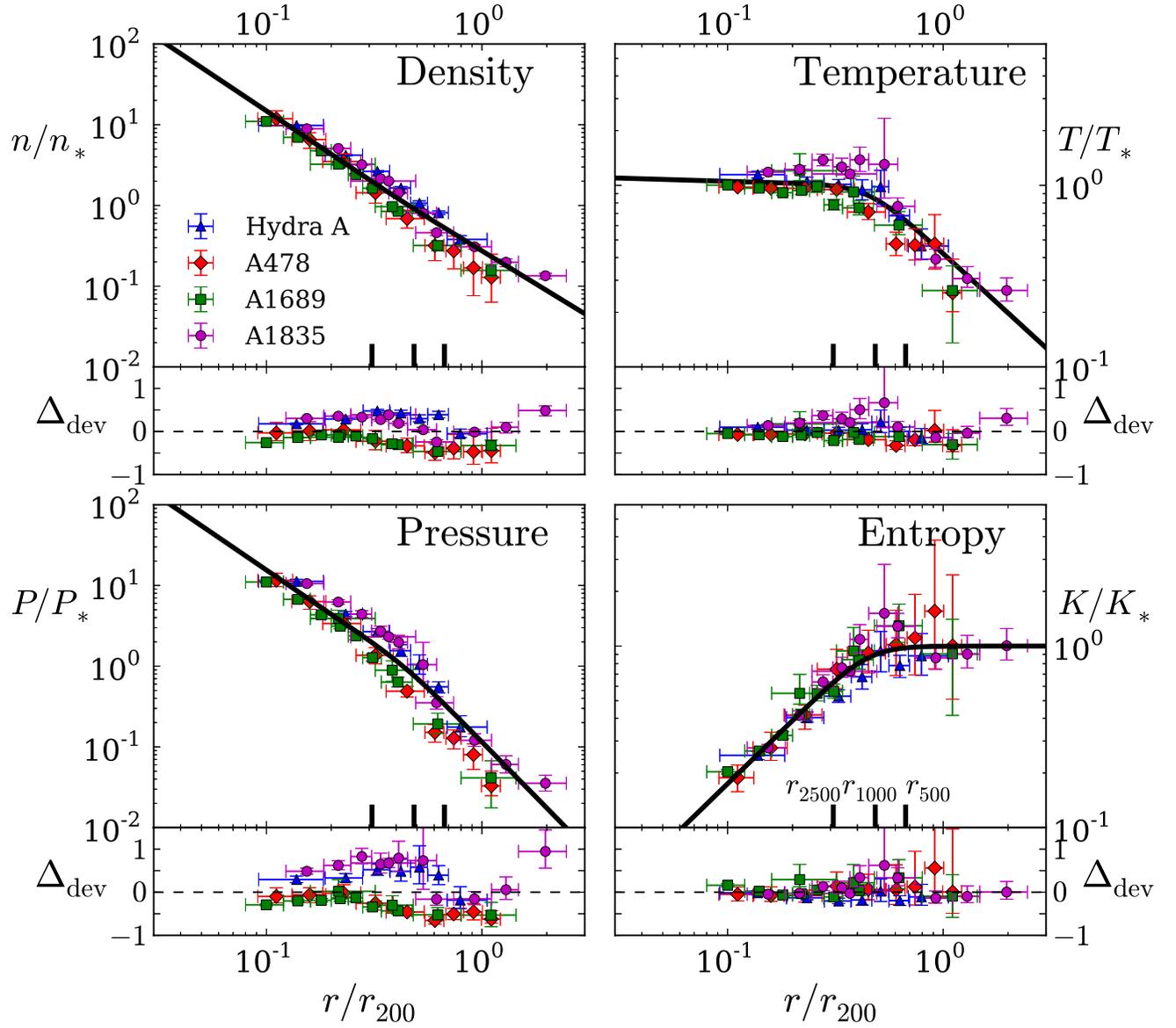}
\caption{Normalized radial profiles of gas number density (top-left), temperature (top-right), pressure (bottom-left) and entropy
 (bottom-right), obtained from joint analysis of the observed gas
 density and temperature profiles.
Red diamonds, blue triangle, green squares and magenta circles represent 
the normalized data for A478, Hydra A, A1689, and  A1835, respectively. 
The black-solid curves show the best-fit models.
The lower subpanels show  the deviations from the respective best-fit profiles.
The overall amplitude of deviations is different from cluster to
 cluster, reflecting the intrinsic scatter between the X-ray observables
 and weak-lensing mass.
The deviations are almost constant with radius, indicating that the
 scaled X-ray profiles are fairly universal.
The black vertical bars denote $r_{2500}$,$r_{1000}$, and $r_{500}$, from left to right.
}
\label{fig:all_prof}
\end{figure*}

Since the number density and temperature profiles are equivalent to the entropy and pressure profiles 
 which have been investigated in Sections \ref{subsec:K} and \ref{subsec:P}, 
we shall focus on the results of the intrinsic covariance based on the self-similar solution ($a=b=2/3$). 
For this, we use the full {\it Suzaku}, {\it XMM-Newton} and {\it Chandra} dataset.
The best-fit profiles of gas density and temperature are shown in the
 top panel of Figure \ref{fig:all_prof}. 
The pressure and entropy profiles corresponding to the best-fit
 parameters (Table \ref{tab:best-fit}) are shown in the bottom panel of
 Figure \ref{fig:all_prof}. 
The normalization factors for the gas density, temperature, pressure, and entropy are 
$n_*=n_0E(z)^2\left(M_{200}E(z)\right)^{3(b-a)/5}$,$T_*=T_0\left(M_{200}E(z)\right)^{3b/5+2a/5}$,
$P_*=n_0T_0E(z)^2\left(M_{200}E(z)\right)^{b}$ and
$K_*=T_0n_0^{-2/3}E(z)^{-4/3}\left(M_{200}E(z)\right)^{a}$,
respectively.

The best-fit density profile describes well the observations as shown in
the top-left panel of Figure \ref{fig:all_prof}.
The residual deviation $\Delta_{\rm dev}= n/f_n-1$ from the best-fit profile is
shown in the lower subpanel. 
Compared to A478 and A1689,
the deviations for Hydra A and A1835,
corresponding to the clusters with high gas fractions (Figure
\ref{fig:fgas_ave}), are high but constant with radius $r/r_{200}$,
showing that the observed scaled density profiles follow the universal
trend within intrinsic scatter.
The logarithmic gradient of the gas density slightly changes from 
$d\ln n_e/d \ln r=-3(\alpha+\gamma)/5\sim-1.9$ at $r\ll r_0$ to
$-3\delta/5\sim-1.6$ at $r\gg r_0$.
The gas density slope outside the X-ray cores ($r>0.1r_{200}$)
is thus shallower than the outer asymptotic slope of the NFW density profile ($-3$). 
Similar results were reported in previous studies of {\it Chandra} and {\it XMM-Newton} observations 
\citep[e.g.][]{Vikhlinin06,Zhang08}.
A possible interpretation of the apparent shallow outer slope is that
the degree of gas clumpiness is increased in the cluster outskirts \citep{Nagai11}.

The X-ray temperature profiles are nearly constant at $0.1r_{200}\simlt r\simlt0.5r_{200}$,
which is consistent with previous studies \citep[e.g.][]{Vikhlinin06,Zhang08}.
The logarithmic slope of the temperature drastically changes at
$\sim0.5r_{200}$ from $3\alpha/5-2\gamma/5\sim0$ to
$-2\delta/5\sim-1.1$. 
The deviation $\Delta_{\rm dev}=T/f_T-1$ is constant with radius and close to
zero, showing high similarity of the temperature profiles over a wide
radial range.

For the gas pressure, the logarithmic gradient steepens from
$d\ln{P}/dln{r}=-\gamma\sim-1.8$ to $-\delta\sim-2.7$ over the full
radial range (see Section \ref{subsec:P} for the case of $\Delta=500$).
The deviations $P/f_P-1$ of Hydra A and A1835 are higher than those of
A1689 and A478, which reflects the large deviations in the gas density.

The entropy profile increases with radius as $\propto r^{1.16}$ at
$r\simlt0.5r_{200}$, and then flattens at radii greater than
$\sim 0.5r_{200}$. 
The inner slope is consistent with the gravitational shock heating
model \citep[$\propto r^{1.1}$;][]{Tozzi01}. 
Our results show that the entropy flattening in the outskirts is caused
not by the shallow outer density slope but  
by the steep temperature drop. 
We find the density slope changes with radius only by $\sim +0.3$,
corresponding to a slope change of $\sim -0.2$ for the entropy,
$K\propto n^{-2/3}$.
On the other hand, the steepening of the temperature slope by $\sim-1.1$
significantly affects the entropy slope because $K\propto T$.

Including the intrinsic covariance matrix in our analysis, we can
constrain the intrinsic scatter between the X-ray observables and
weak-lensing masses.
We find the intrinsic scatter of the gas density $\sigma_{\ln n}$ is
greater than that of the temperature $\sigma_{\ln T}$  
for our sample of the clusters. 
This is qualitatively consistent with the trends in the deviation
profiles shown in Figure \ref{fig:all_prof}. 

From Bayesian inference, we obtain a $1\sigma$ lower limit on the correlation
coefficient, $\rho>0.47$.
On the other hand, we find a maximum-likelihood estimate of $\rho=0.96$,
which is high and close to the upper bound ($\rho=1$) of the parameter range.
We thus computed the probability $\mathcal{P}(\geq |\rho |)$
that the correlation coefficient of two random variables for a sample
size of 4 is greater than $0.96$,
finding $\mathcal{P}(\geq |\rho |)=0.04$, which is very small and rules
out the possibility that the high correlation coefficient is randomly generated.

Since the intrinsic scatter of the quantity $X=Tn^p$ is expressed as 
$\sigma_{\ln X}^2=\sigma_{\ln T}^2+p^2\sigma_{\ln n}^2 + 2 p\rho\sigma_{\ln n}\sigma_{\ln T}$,
the observed positive coefficient indicates 
that the third terms for the entropy ($p=-2/3$) and pressure ($p=1$) 
are negative and positive, respectively.
Thus, the intrinsic scatter of the entropy is smaller than that of the pressure,
which is consistent with the results from individual analyses of the
pressure and entropy (Sections \ref{subsec:K} and \ref{subsec:P}).
The positive correlation between the gas density and temperature can
also be seen in their deviation profiles in Figure \ref{fig:all_prof}.
Due to this positive correlation, the deviation amplitude for the
pressure is larger than that for the entropy.
Similarly, a joint X-ray and weak-lensing analysis of \citet{Okabe10c} derived
bivariate $M$-$T$ and $M$-$M_{\rm gas}$ scaling relations for 12
clusters at $\Delta=500$, finding
that the intrinsic scatter between the gas mass and the temperature is
positively correlated.

Theoretical predictions for the intrinsic correlation
between the gas mass and pressure are rather controversial.
Using an adaptive-mesh refinement code, \cite{Kravtsov06} found from their
simulations that the gas-temperature deviations from the $M$-$T$ relation
are anti-correlated with the gas-mass deviations from the $M$-$M_{\rm gas}$ relation. 
On the other hand, \cite{Stanek10} showed  that the temperature and
gas-mass deviations are positively correlated with each other. 
Therefore, a larger sample is required to constrain the intrinsic
correlations between cluster properties, and it will allow us to
investigate the functional form of the radial profile with the lowest
scatter and optimal mass proxies based on the principal component analysis
\citep{Okabe10c}.

Now we examine the validity of the hydrostatic-equilibrium assumption:
\begin{eqnarray}
\frac{1}{\rho_g}\frac{dP_g}{dr}=-\frac{GM}{r^2},
\end{eqnarray}
where $\rho_g$ and $P_g$ are the gas mass density and thermal pressure, respectively.
Since $\rho_g\propto n_e$ and $P_g\propto P_e$, 
the best-fit parameters (Table \ref{tab:best-fit}) in equations (\ref{eq:P}) and (\ref{eq:n}) 
allow us to determine the radius beyond which the enclosed
hydrostatic mass unphysically decreases (i.e., $dM/dr<0$).
We find that the breakdown of the assumption occurs at $r\sim0.84r_{200}\sim1.3r_{500}$,
which is consistent with previous studies \citep[e.g.][]{Mochizuki14,Kawaharada10,Ichikawa13}.
The breakdown of the hydrostatic assumption underestimates hydrostatic
mass estimates at $r\simgt 1.3r_{500}$.
Indeed, Figure \ref{fig:Mratio_ave} shows 
that the hydrostatic masses are on average much lower than the weak-lensing masses at $r>r_{500}$.

The methodology we have applied here, which simultaneously fits X-ray
observable profiles taking into account multivariate scaling relations,
stems from our earlier work on the multivariate scaling relations
\citep{Okabe10c}.
In the traditional method, one obtains scaling relations between the
total mass and X-ray observables at a given reference overdensity (e.g.,
$\Delta=200$ or 500). 
Furthermore, we have also established the average
scaled radial profiles and intrinsic scatters for X-ray observables. 
In contrast to the traditional method, this new method enables us to simultaneously constrain the shape of the
radial profile, normalization, and its intrinsic scatter,
where the normalization of the profile corresponds to the
mass-observable scaling relation.
Importantly, this method does not require an iterative procedure to reconstruct
mass-observable scaling relations \citep{Vikhlinin09a} because the
scaling parameters at specific overdensities are determined by lensing
and hence independent of X-ray observables.

We note that this method is complementary to stacking approaches
\citep{PlanckSZPprof,Okabe13,Umetsu14}, which allow us to derive
ensemble-averaged cluster profiles in a model-independent way.
This new method to measure the characteristic shapes and normalizations
of profiles will provide us a powerful means to establish the mass
proxy for cluster cosmology studies, especially for the forthcoming
large-sky X-ray surveys (e.g. {\it eROSITA}).

\begin{table*}
\caption{Best-fit parameters for the universal entropy and pressure
 function (equations (\ref{eq:K}) and (\ref{eq:P})),
 obtained from a simultaneous fit to the observed electron number
 density, X-ray temperature, and weak-lensing masses.
} \label{tab:best-fit}
\begin{center}
\begin{tabular}{cccccc}
\hline
\hline
      $\log_{10}(n_0)$
     & $T_0$
     & $a$
     & $b$
     & $\alpha$
     & $\beta$ \\
      $\log_{10}(1$cm$^{-3}$)     & [keV]
     & 
     & 
     & 
     &  \\
\hline
            $-3.60_{-0.14}^{+0.15}$
             & $1.27_{-0.19}^{+0.24}$
             & $2/3$ (fixed)
             & $2/3$ (fixed)
             & $1.16_{-0.12}^{+0.17}$
             & $5.52_{-2.64}^{+2.87}$ \\
\hline
      $\gamma$
     & $\delta$
     & $r_0$ 
     & $\sigma_{\ln n}$
     & $\sigma_{\ln T}$
     & $\rho$ \\ 
     &
     & [$r_{200}$] 
     &
     &
     &\\
\hline
              $1.82_{-0.30}^{+0.28}$
             & $2.72_{-0.35}^{+0.34}$
             & $0.45_{-0.07}^{+0.08}$ 
             & $0.22_{-0.04}^{+0.05}$
             & $0.07_{-0.04}^{+0.06}$
             & $>0.49$\\
\hline
\end{tabular}
\end{center}
\end{table*}

\section{Summary} \label{sec:sum}

We have performed a joint X-ray and weak-lensing analysis of a sample of
four relaxed clusters (Hydra A, A478, A1689, and A1835), 
which had been deeply observed to date by both {\it Suzaku} and
Subaru out to virial radii.
We have shown that the X-ray hydrostatic mass estimates are
systematically underestimated, where the average hydrostatic-to-lensing
mass ratio decreases from $\sim 70\%$ at $r_{500}$ to $\sim 30\%$ at
$r_{\rm vir}$.
This radial dependence is detected at the $3\sigma$ significance level.
The average gas mass fraction from weak-lensing mass estimates increases
with radius, and agrees with the cosmic baryon fraction within the
virial radius.
The outskirts entropy is shown to be tightly correlated with the total
cluster mass from lensing, but not with the gas mass fraction within the
virial radii.

We have developed a new advanced method for determining normalized
cluster radial profiles for multiple X-ray observables by simultaneously
taking into account both their radial dependence and multivariate
scaling relations with weak-lensing masses.
This method stems from the techniques for determining the multivariate scaling relations
\citep{Okabe10c} and is complementary to stacking approaches. 
In the paper, we first used
this method to individually reconstruct each universal X-ray observable
function (Sections \ref{subsec:K} and \ref{subsec:P}).  We then applied
it to simultaneously determine two X-ray observable functions by taking
into account their intrinsic covariance (Section \ref{subsec:joint}).
A combination of complementary data sets of the weak-lensing masses and radii and X-ray observables 
is essential to this method.

We find the gas entropy, pressure, and density profiles are all remarkably
 self-similar when scaled with the weak-lensing $M_{200}$ mass and $r_{200}$ radius. 
The entropy monotonically increases out to $\sim 0.5r_{200}\sim r_{1000}$ 
following the accretion shock heating model $K(r)\propto r^{1.1}$ \citep{Tozzi01},
and flattens at $\simgt 0.5r_{200}$.  The logarithmic gradient of the
gas density becomes slightly shallower at $r\sim 0.5r_{200}$.  A
possible interpretation for this is that the degree of gas clumpiness is
increased in the outskirts.
The temperature profile is constant at $0.1r_{200}<r<\simlt
0.5r_{200}$, and sharply drops off outside $\sim 0.5 r_{200}$. 
The bivariate scaling functions of the gas density and temperature reveal 
that the flatness of the outskirts entropy profile is caused by the steepening of
 the temperature, rather than the flattening of the gas density.
Thus, gas clumpiness alone cannot be responsible for all of the flatness of the outskirts entropy.
The pressure profile exhibits a steep outer slope,
in good agreement with the averaged {\it Planck} Sunyaev-Zel'dovich pressure profile. 
The assumption of hydrostatic equilibrium breaks down beyond $\sim0.84r_{200}\sim1.3r_{500}$.

Our cluster sample may not be representative of a homogeneous class of
actual clusters because the clusters were selected solely by the current availability of both {\it
Suzaku} and Subaru observations. 
The sample clusters span a range of halo mass, concentration, X-ray luminosity and redshift.
Nevertheless, we find the universality of the scaled entropy profiles.
This indicates that the thermalization mechanism in the ICM over the
entire region ($>0.1r_{200}$) is controlled by gravitation in a common
way for all clusters.
The entropy flattening in cluster outskirts appears to
be a common phenomena, not limited to a special class of clusters.
This demonstrates that the heating efficiency in the outskirts
 needs to be modified from the standard $r^{1.1}$ law.
Since the current sample is small and limited by the availability of data,
a further systematic multi-wavelength study of cluster outskirts is
vitally important to understand the physical state of the ICM.

\section*{Acknowledgments}

This work was supported by World Premier International Research Center
Initiative (WPI Initiative), MEXT, Japan. 
N. Okabe (26800097), M. Takizawa (26400218), and K. Sato (25800112) are supported by a
Grant-in-Aid from the Ministry of Education, Culture, Sports, Science,
and Technology of Japan.
K. Umetsu acknowledges partial support from the National Science Council of
Taiwan (grant NSC100-2112-M-001-008-MY3).
Y. -Y. Zhang acknowledges support by the German BMWi through
the Verbundforschung under grant 50\,OR\,1304.

\bibliographystyle{apj}
\bibliography{my,hydraa}

\appendix

\label{lastpage}

\end{document}